\documentclass[3p]{elsarticle}
\biboptions{number}

\usepackage{bm}
\usepackage{hyperref}
\usepackage{epsfig}
\usepackage{graphicx,epsf}
\usepackage{color}
\usepackage{amssymb}
\usepackage{amsopn}
\usepackage{amsmath}
\def\be{\begin{equation}}
\def\ee{\end{equation}}
\def\bea{\begin{eqnarray}}
\def\eea{\end{eqnarray}}
\def\p{\partial}

\def\vp{{{\varphi}}}
\def\dvp{{\delta\varphi}}

\newcommand\dvpK[1]{{\delta\varphi_{{#1}K}}}

\def\cs2{c_{\rm{s}}^2}
\def\wt{\widetilde}
\def\dij{\delta_{ij}}
\newcommand\eq[1]{Eq.~(\ref{#1})}
\newcommand\eqs[1]{Eqs.~(\ref{#1})}
\def \beg {\begin{enumerate}}
\def \en {\end{enumerate}}

\def\vpb{\varphi_0}
\def\Pb{P_0}
\def\rhob{\rho_0}
\def\Xb{X_0}

\def\drho{{\delta\rho}}
\def\dP{{\delta P}}
\def\dPn{{\delta P_{\rm{nad}}}}
\def\dX{\delta X}

\def\cph{c_{\rm{ph}}^2}
\def\cs{c_{\rm{s}}^2}

\begin{document}

\title{The non-adiabatic pressure in general scalar field systems}
\author[qm]{Adam J.~Christopherson\corref{cor1}}
\ead{a.christopherson@qmul.ac.uk}
\author[qm]{Karim A.~Malik}
\ead{k.malik@qmul.ac.uk}

\address[qm]{Astronomy Unit, School of Mathematical Sciences, \\
Queen Mary University of London, \\
Mile End Road, \\
London, E1 4NS, \\
United Kingdom}

\cortext[cor1]{Corresponding author}
\date{\today}
\begin{abstract}

We discuss the non-adiabatic or entropy perturbation, which controls
the evolution of the curvature perturbation in the uniform density
gauge, for a scalar field system minimally coupled to gravity with
non-canonical action. We highlight the differences between the sound
and the phase speed in these systems, and
show that the non-adiabatic pressure perturbation vanishes in the
single field case, resulting in the conservation of the curvature
perturbation on large scales.
\end{abstract}
\begin{keyword}
Scalar fields in cosmology \sep non-adiabatic pressure \sep sound speed
\PACS 98.80.Cq \sep 98.80.Jk \hfill Phys. Lett. B 675 (2009) 159-163,  arXiv:0809.3518v3
\end{keyword}

\maketitle

\section {Introduction} 
\label{sec:intro}

Providing a mechanism to generate the primordial spectrum of density
fluctuations that later on in the history of the universe source the
Cosmic Microwave Background anisotropies and the Large Scale Structure
is one of the strongest points in favour of inflation
\cite{LLBook,Komatsu:2008hk}.
The calculation of this spectrum is a crucial tool in studying the early universe 
and inflation.
Instead of following the evolution of the field perturbations directly
it is easier to relate them to conserved quantities which are constant
in time on super-horizon scales, such as the curvature perturbation on
uniform density hypersurfaces, $\zeta$. This allows us to side-step
unresolved issues of inflationary cosmology such as the decay of the
scalar field(s) driving inflation into the standard model particles,
and directly relate the epoch when the primordial density spectrum is
generated to the epoch of structure formation after the perturbations
reenter the horizon.

Conserved quantities, such as $\zeta$, will only be constant on
super-horizon scales and in the very simplest cases such as slow-roll
single field inflation, and will evolve in others such as multi-field
or non slow-roll inflation.
The evolution equation for the curvature perturbation $\zeta$ follows
directly from the divergence of the energy-momentum tensor
\cite{WMLL}, and can be written on large scales as
\be
\label{eq:zetadotlarge}
\dot{\zeta}=-\frac{H}{(\rhob+\Pb)} \,\dPn\,,
\ee
where $\dPn$ is the non-adiabatic pressure
perturbation which we shall study in detail below.
Hence $\zeta$ will be constant if $\dPn=0$. On the other hand, the
non-adiabatic pressure can be used to generate or amplify the
curvature perturbation, which is exploited in the curvaton mechanism
\cite{curvaton}.

There has been a lot of recent interest in scalar fields with
non-canonical actions. Early work in the realm of the early universe
included that of k-inflation
\cite{ArmendarizPicon:1999rj,Garriga:1999vw}, and more current work
includes that of the string theory motivated Dirac-Born-Infeld (DBI)
inflation
\cite{Silverstein:2003hf,Alishahiha:2004eh,Seery:2005wm,Lidsey:2007gq,
Huston:2008ku}. Scalar fields with non-canonical actions have also
recently been considered as dark energy candidates (see,
e.g.~Ref.~\cite{Unnikrishnan:2008ki}).

We study here if and under what conditions the curvature perturbation
$\zeta$, remains constant on large, super-horizon scales, for a scalar
field system derived from a general, non-canonical action. We focus on
the classical aspects of the system, which allows us to treat the
scalar fields as a non-standard perfect fluid. For aspects of
quantisation and a more field-centred view see
Ref.~\cite{Langlois}.

The Letter is organised as follows: in the next section we study
thermodynamic relations between the pressure, energy density and
entropy of the system. This is followed by a discussion regarding the
adiabatic sound speed and phase speed of a system. In Section
\ref{sec:equations} we give the governing equations for the
non-canonical scalar field system, both in the background and to first
order in the perturbations and describe specific examples of adiabatic
sound speed and phase speeds.  In Section \ref{sec:conditions} we present
conditions for the non-adiabatic pressure perturbation to vanish,
and we conclude the Letter with a short discussion in Section
\ref{sec:conclusions}.

Throughout this Letter we assume a spatially flat
Friedmann-Robertson-Walker (FRW) background spacetime and use
coordinate time, $t$. Derivatives with respect to coordinate time are
denoted by a dot. Greek indices, $\mu,\nu,\lambda$, run from
$0,\ldots,3$, while lower-case Latin indices, $i,j,k$, take the value
$1,2$, or $3$.

\section{Thermodynamic relations}
\label{sec:thermo}

Considerable physical insight can be gained by studying the
thermodynamic properties of a system. In order to keep the discussion
as simple as possible, we focus here on a single fluid system. For an
analysis including an arbitrary number of interacting fluids and
canonical fields see e.g.~Ref.~\cite{MW2004}.

\subsection{Entropy and non-adiabatic pressure}
\label{sec:entropy}

A general thermodynamic system is fully characterised by three
variables, of which only two are independent. Here we choose the
energy density $\rho$, and the entropy $S$, as independent variables,
with the pressure $P$ being $P\equiv P(\rho,S)$. The pressure
perturbation can then be expanded into a Taylor series as
\be
\label{eq:dP}
\dP = \frac{\p P}{\p S}\delta S 
+ \frac{\p P}{\p \rho}\drho \,.
\ee
This can be cast in the more familiar form 
\bea
\label{defdPnad}
\dP=\dPn+\cs\delta\rho\,,
\eea
by introducing the adiabatic sound speed
\bea
\label{defcs2}
\cs\equiv\left.\frac{\p P}{\p \rho}\right|_{S}\,,
\eea
and by identifying the non-adiabatic pressure perturbation as
$\dPn\equiv\left.\frac{\p P}{\p S}\right|_{\rho}\delta S$
\cite{WMLL}.  
Equations (\ref{eq:dP}) and (\ref{defdPnad}) provide an intuitive and
convenient definition for the adiabatic and the entropy, or
non-adiabatic, perturbations in our system: the single adiabatic
degree of freedom is proportional to the energy density, all other
degrees of freedom will contribute to the non-adiabatic perturbation.
This definition can easily be extended to more than two degrees of
freedom as we shall show below, and can also be extended to higher
order in the perturbations, as we shall briefly demonstrate at the end
of this section.

With the above definitions, we see from \eq{defdPnad} that the adiabatic
sound speed $\cs$ is of zeroth order in the perturbations, and since
all background quantities depend only on time we get from \eq{defcs2}
\be
\label{cs2back}
\cs=\frac{\dot P_0}{\dot\rho_0}\,,
\ee
where subscript ``0'' denotes a background quantity.
Together with the behaviour of $\delta P$ and $\delta\rho$ under a
gauge-transformation $\tilde t= t-\delta t$, namely $\wt{\delta\rho}
=\delta\rho+\dot\rho_0\delta t$ and $\wt{\delta P} =\delta P+\dot
P_0\delta t$, see e.g.~Ref.~\cite{MM2008}, only the definition for the
sound speed $\cs$ given in \eq{cs2back} renders the non-adiabatic
pressure perturbation $\dPn$ gauge-invariant.

We now focus on a system with two degrees of freedom. In order to
obtain the non-adiabatic pressure perturbation in terms of different
variables, we simply change from $\rho$ and $P$ to the new variables
denoted here $\vp$ and $X$, which we shall identify in the following
section with the amplitude of the scalar field and its kinetic term,
respectively.
The energy density and the pressure are then functions of $\vp$ and
$X$, that is $P=P(\vp,X)$ and $\rho=\rho(\vp,X)$, and
we can therefore write
\bea
\label{eq:dP1}
\dP = \frac{\p P}{\p \vp}\dvp +\frac{\p P}{\p X}\delta X \,,
\eea
and
\bea
\label{eq:drho}
\drho = \frac{\p \rho}{\p \vp}\dvp+\frac{\p \rho}{\p X}\delta X \,.
\eea
Substituting \eqs{eq:dP1} and (\ref{eq:drho}) into \eq{eq:dP} we
obtain
\be 
\label{eq:nad}
\delta P_{\rm{nad}}
=\rho_{,\vp}
\left(\frac{ P_{,\vp}}{\rho_{,\vp}}-\cs\right)\dvp{}
+\rho_{,X}\left(\frac{P_{,X}}{\rho_{,X}}-\cs\right)\delta X\,,
\ee
where, for example, $P_{,\vp}\equiv\p P/\p \vp$.
Note that in \eq{eq:nad} the terms in brackets are of zeroth order,
and hence $\delta P_{\rm{nad}}$ can be evaluated once $P(\vp,X)$ has
been specified using background quantities only, which we shall do
below in Section \ref{sec:conditions}.

Equation (\ref{eq:nad}) can readily be extended to more than two
degrees of freedom.  For example, if the energy density and the
pressure are functions of $N$ fields $\vp_I$ and $X$, that is
$P=P(\vp_I,X)$ and $\rho=\rho(\vp_I,X)$, then
\be 
\label{eq:nad:mult}
\delta P_{\rm{nad}}
=\sum_K\rho_{,{\vp_K}}
\left(\frac{ P_{,{\vp_K}}}{\rho_{,{\vp_K}}}-\cs\right)\dvpK{}
+\rho_{,X}\left(\frac{P_{,X}}{\rho_{,X}}-\cs\right)\delta X\,.
\ee

To extend Eq.~(\ref{eq:dP}) to higher order we simply do not
truncate the Taylor expansion at the linear order, that is
\bea
\label{eq:dP2}
\dP = \frac{\p P}{\p S}\delta S 
+ \frac{\p P}{\p \rho}\drho  +\frac{1}{2}\Bigg[
\frac{\p^2 P}{\p S^2}\delta S^2 
+\frac{\p^2 P}{\p \rho\p S}\drho \delta S 
+\frac{\p^2 P}{\p \rho^2}\drho^2\Bigg]+\ldots\,.
\eea
The entropy, or non-adiabatic pressure perturbation at second order,
for example, is then found from \eq{eq:dP2}, and expanding $\delta
P\equiv\delta P_1+\frac{1}{2}\delta P_2$, and similarly $\delta\rho$,
we get \cite{Malik:2003mv}
\be
\delta P_{2\rm{nad}}
=\delta P_2-\cs\delta\rho_1-\frac{\p \cs}{\p\rho}\ \delta\rho_1^2\,.
\ee
After this higher order excursion, we return to linear theory below.

\subsection{Sound speed and phase speed}
\label{sec:soundphase}

Many oscillating systems can be described by a wave equation, that is
a second order evolution equation of the form
\be
\label{eq:wave_equ}
\frac{1}{\cph}\ddot\phi-\nabla^2\phi+F(\phi,\dot\phi)=0\,,
\ee
where $\phi$ is, for example, the velocity potential or the scalar field
amplitude, $F(\phi,\dot\phi)$ is the damping term, and $\cph$ is the
phase speed, i.e.~the speed with which perturbations travel through
the system \cite{arnold}.

There is some confusion in the literature on the meaning of adiabatic
sound speed and phase speed. Although this seems not to affect the
results of previous works, and the adiabatic sound speed is often
simply used as a convenient shorthand for $\dot{\Pb}/\dot{\rhob}$, as
defined in \eq{cs2back}, it is often confusingly used to denote the
phase speed. We take the opportunity to clarify these issues here.

The adiabatic sound speed defined above describes the response
of the pressure to a change in density at constant entropy
\footnote{More intuitive is the introduction of the adiabatic sound
speed using the compressibility,
$\beta\equiv\frac{1}{\rho}\frac{\p\rho}{\p P}\Big|_{S}
=\frac{1}{\rho\cs}$, which describes the change in density due to a
change in pressure while keeping the entropy constant.}, 
and is the speed with which pressure perturbations travels through a
classical fluid. A collection of scalar fields can also be described
as fluid, but the analogy is not exact. 
Whereas in the fluid case phase speed $\cph$ and adiabatic sound speed
$\cs$ are equal, this is not true in the scalar field case and the
speed with which perturbations travel is given \emph{only} by $\cph$.
We shall illustrate the difference in $\cph$ and $\cs$ for a concrete
example in Section \ref{sec:speeds} below.

\section {Governing equations}
\label{sec:equations}

We now derive the governing equations for a scalar field system with
general non-canonical action. Consider a general Lagrangian for a
single scalar field $\vp$,
\be
\label{eq:lagrangian}
\mathcal{L}=f(X,\vp)\,, 
\ee
where
$X=-\frac{1}{2}g^{\mu\nu}\vp_{,\mu}\vp_{,\nu}$.

The energy-momentum tensor is defined as \cite{LLBook}
\be 
\label{eq:energymomentum}
T_{\mu\nu}=-2\frac{ \p \mathcal{L}}
{\p g^{\mu\nu}}+g_{\mu\nu}\mathcal{L} \,, 
\ee
and substituting the Lagrangian Eq.~(\ref{eq:lagrangian}) into 
Eq.~(\ref{eq:energymomentum}) we get
\be
\label{eq:EM2}
{T^{\mu}}_{\nu}
=f_{,X}g^{\mu\lambda}\vp_{,\lambda}\vp_{,\nu}+{\delta^\mu}_\nu f\,. 
\ee
The energy-momentum tensor for a perfect fluid is
\be 
\label{eq:EMfluid}
{T^\mu}_\nu=(\rho+P)u^\mu u_\nu+P{\delta^\mu}_\nu\,,
\ee
where $P$ and $\rho$ are the pressure and energy density,
respectively, and $u^\mu$ is the four-velocity, subject to the
constraint $u^\mu u_\mu=-1$.

We consider here only scalar perturbations in a flat FRW background
with line element \cite{Bardeen80,KS,MFB}
\bea
\label{eq:metric}
ds^2=-(1+2A)dt^2+2aB_{,i}dx^idt +a^2\left[(1-2\psi)\dij+2E_{,ij}\right]dx^idx^j\,,
\eea
where $a=a(t)$ is the scale factor, $A$ is the lapse function, $B$ and
$E$ make up the shear ($\sigma=a^2\dot{E}-aB$), and $\psi$ is the
dimensionless curvature perturbation.
The four-velocity is then given by
\be
\label{eq:4velocity}
u^{\mu}=\left[(1-A),\frac{1}{a} v_,^i\right]\,,
\ee
where $v$ is the scalar velocity potential of the fluid.

We split quantities into a time-dependent background and a time- and
space-dependent perturbation, and get e.g.~for the scalar field
$\vp=\vpb(t)+\dvp(t,x^i)$.
The energy density and pressure of the scalar field are then, in the
background
\be
\label{rho_back}
\rhob=2f_{,X}\Xb-f\,, \qquad \Pb=f\,,
\ee
where $\Xb=\frac{1}{2}\dot{\vpb}^2$, and to first order in the 
perturbations 
\bea
\label{eq:perturbed}
\drho&=&\left(f_{,X}+2\Xb f_{,XX}\right)\delta X
+\left(2\Xb f_{,X\vp}-f_{,\vp}\right)\dvp{}\,, \nonumber \\ 
\dP&=&f_{,X}\delta X+f_{,\vp}\dvp{}\,,
\eea
where we have defined $\dX=\dot{\vpb}\dot{\dvp}-A\dot{\vpb}^2$. It is
convenient to work with the covariant velocity perturbation,
$V=a(v+B)$, which in terms of the scalar field is given from ${T^i}_0$
component of the energy-momentum tensor as
\be
\label{Vrel}
V=-\frac{\dvp}{\dot{\vpb}}\,.
\ee
%

\subsection{Background}
\label{sec:background}

The conservation of energy-momentum is ${T^\mu}_{\nu;\mu}=0$, which in
the background reduces to
\be 
\label{eq:bkgd}
\dot{\rhob}=-3H(\rhob+\Pb)\,.
\ee
In terms of the scalar field, this becomes the evolution equation
\be
\label{eq:bkgdfield}
\ddot{\vpb}\left(f_{,XX}\dot{\vpb}^2+f_{,X}\right)+f_{,X\vp}\dot{\vpb}^2
-f_{,\vp}+3Hf_{,X}\dot{\vpb}=0 \,.
\ee
The Friedmann constraint equation is given by
\be
H^2=\frac{8\pi G}{3}\rhob\,,
\ee
where $H\equiv\dot a/a$ is the Hubble parameter.

\subsection{First order}
\label{sec:firstorder}

Conservation of energy-momentum to linear order gives the perturbed
energy conservation equation
\be 
\label{eq:fluidevolution}
\dot{\drho}=-3H(\drho+\dP)+(\rho_0+P_0)\left[3\dot{\psi}
-\frac{\nabla^2}{a^2}(\sigma+V)\right]  \,.
\ee
This can be readily rewritten in terms of the curvature perturbation
on uniform density hypersurfaces $\zeta\equiv-\psi-
H\delta\rho/\dot\rho$, and working in the uniform density gauge, where
$\delta\rho\equiv0$ and hence $\dP=\dPn$ and $\zeta\equiv -\psi$, we
get \cite{WMLL}
\be
\label{eq:zetadot}
\dot{\zeta}=\frac{-H}{(\rhob+\Pb)}\dPn-\frac{\nabla^2}{3a^2}(\sigma+V) \,.
\ee
Equation (\ref{eq:zetadot}) reduces on large scales, where gradient
terms can be neglected, to \eq{eq:zetadotlarge} and shows that if the
pressure perturbation is adiabatic, then the curvature perturbation
is conserved. See Ref.~\cite{WMLL} for a more detailed discussion.

In the following we will also need the constraint equation (see
e.g.~Ref.~\cite{MW2004}),
\be
\label{deltarho_const}
\delta\rho-3H(\rhob+\Pb)V+\frac{H}{4\pi G a^2}k^2\sigma=0\,,
\ee
where we have replaced spatial Laplacians with the wave-numbers of
their respective eigenvalues, that is $\nabla^2\rightarrow -k^2,$ and
have chosen the flat gauge, where $\psi=0=E$. On large scales, where
gradient terms can be neglected, we can rewrite \eq{deltarho_const} in
terms of the scalar field, using \eqs{eq:perturbed} and (\ref{Vrel}),
as
\be
\label{X_vp_rel}
\left(f_{,X}+\dot{\vpb}^2 f_{,XX}\right)\dX
+\left(\dot{\vpb}^2 f_{,X\vp}-f_{,\vp}
+3Hf_{,X}\dot{\vpb}\right)\dvp{}=0 \,,
\ee
giving a convenient relation between $\dX$ and $\dvp$.

\subsection {Speeds}
\label{sec:speeds}

The scalar field system we are considering here can be described by
``borrowing'' terminology from fluid dynamics, though the analogy is
not exact. If we are interested in the speed with which perturbations
travel through the system, we have to calculate the phase speed, which
can be read off from the perturbed Klein-Gordon equation governing the
evolution of the scalar field. This is just a damped wave equation,
like \eq{eq:wave_equ}, and so by comparing the coefficients of the
second order temporal, and second order spatial derivatives, we obtain
the phase speed.

For the general non-canonical Lagrangian, (\ref{eq:lagrangian}), the
evolution equation for the first order scalar field perturbation is
found by substituting Eq.~(\ref{eq:perturbed}) into
(\ref{eq:fluidevolution}), and is given here already in the form of
Eq.~(\ref{eq:wave_equ}) as
\be
\label{eq:pertevolution}
\frac{1}{\cph}\ddot{\dvp}+\Big[\frac{3H}{\cph}+C_1\Big]\dot{\dvp}
+\Big[\frac{k^2}{a^2}+C_2\Big]\dvp=0\,,
\ee
where the coefficients $C_1$ and $C_2$, both functions of $\vp$ and
$X$, are given in Eq.~(\ref{eq:pertevolution2}) in the Appendix. 
We can therefore read off the phase speed as
\be
\label{eq:cphgeneral}
\cph=\frac{f_{,X}}{f_{,X}+2\Xb f_{,XX}}\,.
\ee 
Using \eq{rho_back}, this reduces to \cite{Garriga:1999vw}
\be
\label{def_cph_final}
\cph=\frac{P_{0,X}}{\rho_{0,X}}\,.
\ee
The adiabatic sound speed for the Lagrangian (\ref{eq:lagrangian}) is
given by
\be
\label{eq:csgeneral}
\cs=\frac{f_{,X}\ddot{\vpb}+f_{,\vp}}
{f_{,X}\ddot{\vpb}-f_{,\vp}+f_{,XX}\dot{\vpb}^2\ddot{\vpb}+
f_{,X\vp}\dot{\vpb}^2} \,.
\ee

The Lagrangian for a canonical scalar field is obtained by setting
$f(X,\vp)=X-U(\vp)$. In this case the adiabatic sound speed reduces to
\be
\label{eq:csstandard}
\cs=1+\frac{2U_{,\vp}}{3H\dot{\vpb}} \,,
\ee
and becomes $\cs=-1$ in slow-roll. The phase speed for a canonical
scalar field is $\cph=1.$

\section{Non-adiabatic pressure for a non-canonical scalar field}
\label{sec:conditions}

We now return to the question of under what conditions the non-adiabatic
pressure in a general scalar field system vanishes.
The non-adiabatic pressure $\dPn$ for the system (\ref{eq:lagrangian})
is found by substituting the expressions for the energy and the
pressure, \eq{rho_back}, into Eq.~(\ref{eq:nad}),
\bea
\label{Pnad_gen}
\dPn&=&\Big[f_{,\vp}\left(1+\cs\right)-2\cs \Xb f_{,X\vp}
\Big]\dvp{}+\Big[f_{,X}\left(1-\cs\right)-2\cs \Xb f_{,XX}
\Big]\dX\,,
\eea
where $\cs$ was given in \eq{eq:csgeneral}.

Using the constraint equation (\ref{X_vp_rel}) and the evolution
equation for the background field, Eq. (\ref{eq:bkgdfield}), we obtain
a simple relationship between $\dX$ and $\dvp$, namely,
\be
\dX=\ddot{\vpb}\dvp \,.
\ee
Substituting this into the expression for the non-adiabatic pressure
perturbation, and using $\cs$, we obtain,
\be
\dPn=0 \,.
\ee
Therefore, the non-adiabatic pressure perturbation $\dPn$ for the
general action (\ref{eq:lagrangian}) vanishes in the large scale limit
without the need to impose any further conditions, just as in the
canonical case \cite{Gordon:2000hv}.
This is in agreement with Ref.~\cite{Langlois}\footnote{We thank
S\'ebastien Renaux-Petel for bringing this to our attention
.}.

This is at first glance a surprising result, since one might assume
that the system has two degrees of freedom, $\dvp$ and $\dX$,
and that one would need to impose slow-roll to ``switch off'' a degree of
freedom in order to obtain zero non-adiabatic pressure.

Yet, in the super-horizon limit $k\to 0$, such a condition is not
necessary since the constraint equation (\ref{X_vp_rel}) eliminates
one degree of freedom.

For multiple scalar fields, however, this equation does not hold, and
hence $\dPn$ is, in general, non-zero. \\

If we study specific models, for example the DBI inflation models of
Refs.~\cite{Lidsey:2007gq,Huston:2008ku}, and loosen the
super-horizon limit $k\to 0$ by instead studying
small but non-zero wavenumbers $k$,
we can calculate the possible small deviation from zero of
the non-adiabatic pressure or entropy perturbation. This is an additional 
observational constraint which can be imposed on
the models, and which is already
strongly constrained by the data \cite{Komatsu:2008hk}. 
Since these issues are outside the aim of this work, we will
discuss them in detail in a forthcoming article \cite{CHM}.

\section{Discussion and conclusions}
\label{sec:conclusions}

We have studied whether the curvature perturbation in a system with a
scalar field minimally coupled to gravity with non-canonical action is
constant on super-horizon scales. Deriving the adiabatic and the
non-adiabatic or entropy perturbation we found that imposing the
strict super-horizon limit $k \to 0$
is sufficient to guarantee conservation of $\zeta$
and hence of the primordial power spectrum, just as in the canonical
case.

The non-adiabatic pressure, or entropy, perturbation is in general
non-zero, if there is more than one degree of freedom in the system,
as is the case in multi-field systems. In this case the curvature
perturbation evolves, even on large scales, and it is necessary to
solve the evolution equations of the scalar fields to construct the
non-adiabatic pressure. However, it is then more efficient to
construct $\zeta$ directly using its definition and substituting in
the energy densities in terms of the fields, instead of first
constructing $\dPn$ and then solving the evolution equation for
$\zeta$, \eq{eq:zetadot}, see \cite{Gordon:2000hv,Langlois}.

We have also highlighted the difference between the phase speed and
the adiabatic sound speed of a system in Section
\ref{sec:speeds}. Although these are the same in classical fluid
systems, they are different in the scalar field models studied here,
with only the phase speed describing the speed with which
perturbations travel.
%
%
We emphasise that only the definition for the adiabatic sound speed,
\eq{cs2back}, enters the gauge-transformation of the pressure
perturbation, as pointed out in Section \ref{sec:entropy}. Similarly,
in the adiabatic case when $\dPn=0$, \eq{defdPnad} reduces to
$\dP=\cs\delta\rho$. Again, this convenient relation between the
pressure and the energy density perturbation is only correct when
using the adiabatic sound speed, as defined in \eq{cs2back}.

\section*{Acknowledgments}

The authors are grateful to Ian Huston, Jim Lidsey, S\'ebastien
Renaux-Petel, and David Seery for useful discussions and comments. AJC
is supported by the Science and Technology Facilities Council (STFC).


\section*{Appendix}
\label{sec:appendix}

The coefficients $C_1$ and $C_2$ in the perturbed evolution equation
for the scalar field, (\ref{eq:pertevolution}), are given by
\bea
\label{eq:pertevolution2}
C_1&=&\frac{\cph}{f_{,X}^2}\left[f_{,\vp}-3Hf_{,X}\dot{\vpb}-
f_{,X\vp}\dot{\vpb}^2\right]\left[3f_{,XX}\dot{\vpb}+\dot{\vpb}^3f_{,XXX}\right] 
+\frac{1}{f_{,X}}\left[\dot{\vpb}f_{,X\vp}+\dot{\vpb}^3f_{,XX\vp}\right]\,,\nonumber\\
C_2&=&\frac{\cph}{f_{,X}^2}\left[f_{,\vp}-3Hf_{,X}\dot{\vpb}
-f_{,X\vp}\dot{\vpb}^2\right]\left[f_{,X\vp}+\dot{\vpb}^2f_{,XX\vp}
-\frac{4\pi G\dot{\vpb}f_{,X}}{H}\left(5\dot{\vpb}^2f_{,XX}+\dot{\vpb}^4f_{,XXX}
+2f_{,X}\right)   \right] \nonumber\\
&&-\frac{4\pi G\dot{\vpb}}{H\cph}\left[3H\dot{\vpb}f_{,X}+f_{,\vp}
-\frac{f_{,X}\dot{\vpb}}{H}(3H^2+2\dot{H})+\cph(\dot{\vpb}^4f_{,XX\vp}+f_{,\vp})\right]\nonumber\\
&&+\frac{1}{f_{,X}}\left[\dot{\vpb}^2f_{,X\vp\vp}-f_{,\vp\vp}+3H\dot{\vpb}f_{,X\vp}\right]
\eea
where we have made use of the following expressions relating metric
perturbations to perturbations in the scalar field derived from
Ref.~\cite{MW2004},
\bea
A&=&\frac{4\pi G}{H}\dot{\vpb}f_{,X}\dvp \,,  \\
\dot{A}&=&\frac{4\pi G}{H^2}\left[Hf_{,\vp}-4\pi G\dot{\vpb}^3f_{,X}^2
-\dot{\vpb}f_{,X}(3H^2+2\dot{H})\right]\dvp 
+\frac{4\pi G}{H}\dot{\vpb}f_{,X}\dot{\dvp} \,, 
\eea
and \eq{deltarho_const}.




\end{document}